\documentclass{article}%
\usepackage{amsmath}
\usepackage{graphicx}
\usepackage{amsfonts}
\usepackage{amssymb}%
\newtheorem{theorem}{Theorem}

\newtheorem{example}[theorem]{Example}

\newtheorem{proposition}[theorem]{Proposition}

\newenvironment{proof}[1][Proof]{\noindent\textbf{#1.} }{\ \rule{0.5em}{0.5em}}
\begin{document}

\title{The Maslov Indices of Hamiltonian Periodic Orbits}
\author{Maurice de Gosson\\Blekinge Institute of Technology \\SE 371 79 Karlskrona\\mdg@bth.se
\and Serge de Gosson\\V\"{a}xj\"{o} University (MSI)\\SE 351 95 V\"{a}xj\"{o}\\serge.degosson@msi.vxu.se}
\maketitle

\begin{abstract}
We use the properties of the Leray index to give precise formulas in arbitrary
dimensions for the Maslov index of the monodromy matrix arising in periodic
Hamiltonian systems. We compare our index with other indices appearing in the literature.

\end{abstract}

\section{Introduction}

There have been in the recent years many papers discussing the
\textquotedblleft Maslov index\textquotedblright\ of periodic Hamiltonian
orbits; the papers by Brack and Jain \cite{BJ}, Creagh and Littlejohn
\cite{CL}, Robbins \cite{Robbins}, and Sugita \cite{sugita} are certainly
major advances. One of the main reasons of the interest in the topic comes
from the fact that these indices are play a crucial role in the determination
of the correct phases in Gutzwiller's trace formula \cite{Gutz}, which is one
of the main tools in the study the spectrum of semiclassical systems
associated to non-integrable Hamiltonians.

The purpose of this Letter is to relate that Maslov index to another index, of
cohomological nature, defined by Leray \cite{Leray} in the transversal case,
and extended by the author \cite{JMPA,Wiley} to the general case, and which
has been used in \cite{paselect} in another context. Our approach will
highlight the role played by the index of inertia of triples of Lagrangian
planes, which can be viewed as a Morse index, and completes the Note Added in
Proof in the recent paper \cite{PB} by Pletyukhov and Brack.

\section{Leray and Maslov Indices}

We shortly review the properties of the Leray and Maslov indices that we will
need; for proofs and details see \cite{JMPA,Wiley}. Let $\sigma(z,z^{\prime
})=p\cdot x^{\prime}-p^{\prime}\cdot x$ be the standard symplectic form on
$\mathbb{R}_{z}^{2n}\equiv\mathbb{R}_{x}^{n}\times\mathbb{R}_{p}^{n}$
($z=(x,p)$, $z^{\prime}=(x^{\prime},p^{\prime})$), that is
\[
\sigma(z,z^{\prime})=Jz\cdot z^{\prime}=(z^{\prime})^{T}Jz\text{
\ \ \textrm{with } \ }J=%
\begin{pmatrix}
0 & I\\
-I & 0
\end{pmatrix}
\text{.}%
\]
The corresponding symplectic group is denoted by $Sp(n)$; the unitary group
$U(n,\mathbb{C})$ is identified with a compact subgroup $U(n)$ of $Sp(n)$ by
the isomorphism%
\[
A+iB\longmapsto%
\begin{pmatrix}
A & -B\\
B & A
\end{pmatrix}
\text{.}%
\]
The Lagrangian Grassmannian $\Lambda(n)$ consists of all $n$-dimensional
subspaces $\ell$ of $\mathbb{R}_{z}^{2n}$ on which $\sigma$ vanishes
identically; $U(n)$ (and hence $Sp(n)$) acts transitively on $\Lambda(n)$. Let
$\ell_{p}=0\times\mathbb{R}_{p}^{n}$ be the momentum plane; the
\textquotedblleft Souriau mapping\textquotedblright\
\begin{equation}
\ell=u\ell_{p}\longmapsto uu^{T} \label{souriau}%
\end{equation}
identifies $\Lambda(n)$ with the manifold $W(n,\mathbb{C})\subset
U(n,\mathbb{C})$ of all \textit{symmetric} unitary matrices; the action of
$U(n,\mathbb{C})$ on $\Lambda(n)\equiv W(n,\mathbb{C})$ is then defined by
\[
u\ell=uwu^{T}\text{ \ \textrm{if} \ }\ell\equiv w\text{.}%
\]
A useful result is that if $w\equiv\ell$ and $w^{\prime}\equiv\ell^{\prime}$,
then
\begin{equation}
\operatorname{rank}(w-w^{\prime})=n-\dim\ell\cap\ell^{\prime}\text{.}
\label{rang}%
\end{equation}
The universal coverings of $Sp(n)$, $U(n,\mathbb{C})$ and $\Lambda(n)\equiv
W(n,\mathbb{C})$ are denoted by $Sp_{\infty}(n)$, $U_{\infty}(n,\mathbb{C}) $
and $\Lambda_{\infty}(n)\equiv W_{\infty}(n,\mathbb{C})$. Recall that
$Sp_{\infty}(n)$ is (as is $U_{\infty}(n,\mathbb{C})$) a group; the
composition law is as follows: if $S_{\infty}$ and $S_{\infty}^{\prime}$ are
the homotopy equivalence classes of two symplectic paths $t\longmapsto S_{t}$,
$t\longmapsto S_{t}^{\prime}$, $0\leq t\leq a$ originating at the identity,
then $S_{\infty}S_{\infty}^{\prime}$ is the homotopy class of the path
$t\longmapsto S_{t}S_{t}^{\prime}$. We have the following identifications (see
e.g. \cite{GS}):
\[
U_{\infty}(n,\mathbb{C})=\{(u,\theta):u\in U(n,\mathbb{C}),\det u=e^{i\theta
}\}
\]
and
\[
W_{\infty}(n,\mathbb{C})=\{(w,\theta):w\in W(n,\mathbb{C}),\det w=e^{i\theta
}\}\text{.}%
\]

The Leray index of $(\ell_{\infty},\ell_{\infty}^{\prime})\equiv
(w,\theta;w^{\prime},\theta^{\prime})$ is defined as follows: $(\ast)$ if
$\ell\cap\ell^{\prime}=0$, then $m$ is given by the formula
\begin{equation}
m(\ell_{\infty},\ell_{\infty}^{\prime})=\frac{1}{2\pi}(\theta-\theta^{\prime
}+i\operatorname*{Tr}\operatorname{Log}(-w(w^{\prime})^{-1}))+\frac{n}%
{2}\text{.} \label{sour}%
\end{equation}
Notice that (\ref{rang}) implies that $\det(I-w(w^{\prime})^{-1})\neq0$, hence
$-w(w^{\prime})^{-1}$ has no eigenvalues on the negative half-line. This
allows us to choose for $\operatorname{Log}$ the usual logarithm; $(\ast\ast)$
if $\ell\cap\ell^{\prime}\neq0$ then choose $\ell_{\infty}^{\prime\prime}$
such that $\ell\cap\ell^{\prime\prime}=\ell^{\prime}\cap\ell^{\prime\prime}=0$
and set
\begin{equation}
m(\ell_{\infty},\ell_{\infty}^{\prime})=m(\ell_{\infty},\ell_{\infty}%
^{\prime\prime})-m(\ell_{\infty}^{\prime},\ell_{\infty}^{\prime\prime
})+\operatorname*{Inert}(\ell,\ell^{\prime},\ell^{\prime\prime}) \label{defta}%
\end{equation}
where the \textquotedblleft index of inertia\textquotedblright\ of the triple
$(\ell,\ell^{\prime},\ell^{\prime\prime})$ is defined by:
\begin{equation}
\operatorname*{Inert}(\ell,\ell^{\prime},\ell^{\prime\prime})=\frac{1}{2}%
(\tau(\ell,\ell^{\prime},\ell^{\prime\prime})-\partial\dim(\ell,\ell^{\prime
},\ell^{\prime\prime})+n)\text{.} \label{formula}%
\end{equation}
We will come back to definition (\ref{defta}) in a moment; let us first make a
short detour to explain (\ref{formula}); for details we refer to
\cite{JMPA,Wiley}. First%
\[
\partial\dim(\ell,\ell^{\prime},\ell^{\prime\prime})=\dim(\ell\cap\ell
^{\prime})-\dim(\ell\cap\ell^{\prime\prime})+\dim(\ell^{\prime}\cap
\ell^{\prime\prime})
\]
is the \textquotedblleft coboundary\textquotedblright\ of the
\textquotedblleft cochain\textquotedblright\ $\dim(\ell,\ell^{\prime}%
)\equiv\dim(\ell\cap\ell^{\prime})$; as we will see below it makes
$\operatorname*{Inert}(\ell,\ell^{\prime},\ell^{\prime\prime})$ become an
integer. The term $\tau(\ell,\ell^{\prime},\ell^{\prime\prime})$ is called the
\textquotedblleft signature\textquotedblright\ (or \textquotedblleft Kashiwara
index\textquotedblright) of the triple $(\ell,\ell^{\prime},\ell^{\prime
\prime})$. It is the difference $\tau^{+}-\tau^{-}$ between the number of $>0$
and $<0$ eigenvalues of the quadratic form $\sigma(z,z^{\prime})+\sigma
(z^{\prime},z^{\prime\prime})+$ $\sigma(z^{\prime\prime},z)$ on $\ell
\times\ell^{\prime}\times\ell^{\prime\prime}$. By definition of $Sp(n)$ as the
group of linear automorphisms leaving $\sigma$ invariant it immediately
follows that $\tau(\ell,\ell^{\prime},\ell^{\prime\prime})$ is a symplectic
invariant, that is, $\tau(S\ell,S\ell^{\prime},S\ell^{\prime\prime})=\tau
(\ell,\ell^{\prime},\ell^{\prime\prime})$ for every $S\in Sp(n)$. Since the
symplectic form $\sigma$ is antisymmetric, the signature changes sign when one
swaps any two of the planes in its argument: for instance $\tau(\ell^{\prime
},\ell,\ell^{\prime\prime})=-\tau(\ell,\ell^{\prime},\ell^{\prime\prime})$,
$\tau(\ell^{\prime\prime},\ell^{\prime},\ell)=-\tau(\ell,\ell^{\prime}%
,\ell^{\prime\prime})$, and so on. In particular $\tau(\ell,\ell^{\prime}%
,\ell^{\prime\prime})=0$ if any two of the planes $\ell,\ell^{\prime}%
,\ell^{\prime\prime}$ are identical. A much less trivial property of the
signature is the following: it is a \textquotedblleft
cocycle\textquotedblright, in the sense that $\partial\tau=0$; explicitly:%
\[
\tau(\ell_{1},\ell_{2},\ell_{3})-\tau(\ell_{2},\ell_{3},\ell_{4})+\tau
(\ell_{1},\ell_{3},\ell_{4})-\tau(\ell_{1},\ell_{2},\ell_{4})=0
\]
for all $4$-tuplets $(\ell_{1},\ell_{2},\ell_{3},\ell_{4})$; it is proven
using elementary (but lengthy) arguments of linear algebra (see \cite{Wiley}).
Finally, one shows by studying the kernel of the bilinear form associated with
$\sigma(z,z^{\prime})+\sigma(z^{\prime},z^{\prime\prime})+$ $\sigma
(z^{\prime\prime},z)$ that%
\[
\tau(\ell,\ell^{\prime},\ell^{\prime\prime})\equiv n+\partial\dim(\ell
,\ell^{\prime},\ell^{\prime\prime})\text{ \ }\operatorname{mod}2\text{;}%
\]
this implies that the index of inertia indeed is an integer as claimed. The
following properties of $\operatorname*{Inert}(\ell,\ell^{\prime},\ell
^{\prime\prime})$ immediately follow from those of the signature: it is a
symplectic invariant in the sense that
\begin{equation}
\operatorname*{Inert}(S\ell,S\ell^{\prime},S\ell^{\prime\prime}%
)=\operatorname*{Inert}(\ell,\ell^{\prime},\ell^{\prime\prime})
\label{sympinv}%
\end{equation}
for every $S\in Sp(n)$; it is a cocycle: $\partial\operatorname*{Inert}=0$,
that is
\begin{align}
&  \operatorname*{Inert}(\ell_{1},\ell_{2},\ell_{3})-\operatorname*{Inert}%
(\ell_{2},\ell_{3},\ell_{4})+\operatorname*{Inert}(\ell_{1},\ell_{3},\ell
_{4})\label{cocycle}\\
-\operatorname*{Inert}(\ell_{1},\ell_{2},\ell_{4})  &  =0\text{.}\nonumber
\end{align}
From the antisymmetry of the signature follows that we have the relations%
\begin{align*}
\operatorname*{Inert}(\ell,\ell,\ell^{\prime\prime})  &
=\operatorname*{Inert}(\ell,\ell^{\prime},\ell^{\prime})=0\\
\operatorname*{Inert}(\ell,\ell^{\prime},\ell)  &  =n-\dim\ell\cap\ell
^{\prime}%
\end{align*}
(observe that $\operatorname*{Inert}$ does not inherit the antisymmetry
property of $\tau$; this is due to the presence of the correcting term
$\partial\dim$ in its definition).

Both the signature and the index of inertia are symplectic invariant measures
of the relative positions of Lagrangian planes; let us illustrate this the
simple case $n=1$:

\begin{example}
When $n=1$ the Lagrangian Grassmannian is the set of all lines through the
origin in the phase plane. If the line $\ell^{\prime}$ lies inside the open
angular sector delimited by $\ell$ and $\ell^{\prime\prime}$ (we assume these
lines oriented) then $\tau(\ell,\ell^{\prime},\ell^{\prime\prime})=-1$; if it
lies outside that sector then $\tau(\ell,\ell^{\prime},\ell^{\prime\prime
})=-1$; the signature is zero in the other cases. It follows that
$\operatorname*{Inert}(\ell,\ell^{\prime},\ell^{\prime\prime})$ is $0$ if
$\ell^{\prime}$ is in the closed angular sector delimited by $\ell$ and
$\ell^{\prime\prime}$, and $+1$ otherwise.
\end{example}

Let us now return to the definition (\ref{defta}) of the Leray index. That the
right-hand side of (\ref{defta}) does not depend on the choice of
$\ell_{\infty}^{\prime\prime}$ follows from the fact that
$\operatorname*{Inert}$ is a cocycle: this is proven in detail in \cite{JMPA}.
Moreover (ibid.) the Leray index satisfies the following fundamental cochain
property: we have
\begin{equation}
m(\ell_{\infty},\ell_{\infty}^{\prime})-m(\ell_{\infty},\ell_{\infty}%
^{\prime\prime})+m(\ell_{\infty}^{\prime},\ell_{\infty}^{\prime\prime
})=\operatorname*{Inert}(\ell,\ell^{\prime},\ell^{\prime\prime})
\label{cobord}%
\end{equation}
for all triples $(\ell_{\infty},\ell_{\infty}^{\prime},\ell_{\infty}%
^{\prime\prime})$. Formula (\ref{cobord}) implies in particular that
\begin{equation}
m(\ell_{\infty},\ell_{\infty})=0\text{ \ \textrm{and} \ }m(\ell_{\infty}%
,\ell_{\infty}^{\prime})+m(\ell_{\infty}^{\prime},\ell_{\infty})=n-\dim
\ell\cap\ell^{\prime}\text{.} \label{propel}%
\end{equation}
One also proves (see again \cite{JMPA,Wiley}) that $m$ is in fact the only
function $\Lambda_{\infty}(n)\times\Lambda_{\infty}(n)\longrightarrow
\mathbb{Z}$ satisfying (\ref{cobord}) and such that $m(\ell_{\infty}%
,\ell_{\infty}^{\prime})$ remains constant when $(\ell_{\infty},\ell_{\infty
}^{\prime})$ moves continuously in such a way that $\ell$ and $\ell^{\prime}$
remain transversal (more generally, if $\dim\ell\cap\ell^{\prime}$ does not
change). The index $m$ is a symplectic invariant; more precisely:
\begin{equation}
m(S_{\infty}\ell_{\infty},S_{\infty}\ell_{\infty}^{\prime})=m(\ell_{\infty
},\ell_{\infty}^{\prime})\text{ \ \textrm{for all} \ }S_{\infty}\in
Sp_{\infty}(n) \label{msl}%
\end{equation}
where $(S_{\infty},\ell_{\infty})\longmapsto S_{\infty}\ell_{\infty}$ is the
group action $Sp_{\infty}(n)\times\Lambda_{\infty}(n)\longrightarrow
\Lambda_{\infty}(n)$.

\begin{example}
\label{un}The Souriau mapping (\ref{souriau}) identifies the line $\ell
(\theta):x\cos\theta+p\sin\theta=0$ with the complex number $w=e^{2i\theta}$;
the Leray index is given by
\begin{align*}
m(\ell(\theta),\ell(\theta^{\prime}))  &  =\left[  \frac{\theta-\theta
^{\prime}}{2\pi}\right]  +1\text{ \ \textrm{if} \ }\theta-\theta^{\prime
}\notin\pi\mathbb{Z}\\
m(\ell(\theta),\ell(\theta^{\prime}))  &  =2k\text{ \ \textrm{if} \ }%
\theta-\theta^{\prime}=2k\pi\text{ \ }(k\in\mathbb{Z)}%
\end{align*}
the covering projection $\Lambda_{\infty}(1)\longrightarrow\Lambda(1)$ being
the mapping $\theta\longmapsto e^{2i\theta}$.
\end{example}

Recalling that $\ell_{p}\equiv0\times\mathbb{R}_{p}^{n}$ the standard Maslov
index $\mu$ on $Sp_{\infty}(n)$ is now defined as follows:
\begin{equation}
\mu(S_{\infty})=m(S_{\infty}\ell_{p,\infty},\ell_{p,\infty}) \label{Maslov1}%
\end{equation}
where $\ell_{p,\infty}$ is the homotopy class in $\Lambda(n)$ of the constant
loop through $\ell_{p}$: $\ell_{p,\infty}\equiv(I,0)$. In view of
(\ref{cobord}) and (\ref{msl}) we have the following essential formula giving
the Maslov index of a product:
\begin{equation}
\mu(S_{\infty}S_{\infty}^{\prime})=\mu(S_{\infty})+\mu(S_{\infty}^{\prime
})-\operatorname{Inert}(SS^{\prime}\ell_{p},S\ell_{p},\ell_{p}) \label{mco}%
\end{equation}
for all $S_{\infty},S_{\infty}^{\prime}\in Sp_{\infty}(n)$. Observe that it
immediately follows from (\ref{propel}) that
\begin{equation}
\mu(I_{\infty})=0 \label{propelo}%
\end{equation}
where $I_{\infty}$ is the unit of $Sp_{\infty}(n)$ (\textit{i.e.,} the
homotopy class in $Sp(n)$ of the constant loop through $I$).

\begin{example}
\label{deux}Let $S_{t}$ be the plane rotation with angle $t$, and consider the
path $t\longmapsto S_{t}$, $0\leq t\leq\alpha$. The image of the momentum axis
$\ell_{p}$ by $S_{t}$ is the line $x\cos\alpha+p\sin\alpha=0$. It follows from
Example \ref{un} that $\mu(S_{\alpha,\infty})=[\alpha/2\pi]+1$ if $\alpha
\neq2k\pi$ and $\mu(S_{2k\pi,\infty})=2k$.
\end{example}

The Maslov index of a loop in $U(n)$ through the identity is \textit{twice}
its winding number:

\begin{proposition}
Let $U_{\infty}$ be the homotopy class in $Sp(n)$ of a loop $t\longmapsto
U_{t}$ in $U(n)$ ($0\leq t\leq T$, $U_{0}=U_{T}=I$). If $U_{\infty}\equiv
u_{\infty}=(I,2k\pi)$ for an integer $k$ then
\begin{equation}
\mu(U_{\infty})=2k \label{loop}%
\end{equation}

\end{proposition}

\begin{proof}
Let us set $\ell_{x}=J\ell_{p}$ (it is just the configuration space
$\mathbb{R}_{x}^{n}\times0$), thus $\ell_{x}\equiv-I$. Let $\ell_{x,\infty
}\equiv(-I,n\pi)$. In view of formula (\ref{defta}) defining the Leray index
in the transversal case we have, since $U\ell_{p}=\ell_{p}$:
\begin{align*}
m(U_{\infty}\ell_{p,\infty},\ell_{p,\infty})  &  =m(U_{\infty}\ell_{p,\infty
},\ell_{x,\infty})-m(\ell_{p,\infty},\ell_{x,\infty})+\operatorname*{Inert}%
(\ell_{p},\ell_{p},\ell_{x})\\
&  =m(U_{\infty}\ell_{p,\infty},\ell_{x,\infty})-m(\ell_{p,\infty}%
,\ell_{x,\infty})\text{.}%
\end{align*}
Now $U\ell_{p}\cap\ell_{x}=\ell_{p}\cap\ell_{x}=0$ so that formula
(\ref{sour}) applies:
\[
m(U_{\infty}\ell_{p,\infty},\ell_{x,\infty})=\frac{1}{2\pi}(4k\pi
-n\pi+i\operatorname*{Tr}\operatorname{Log}I)+\frac{n}{2}=2k\text{.}%
\]
Similarly
\[
m(\ell_{p,\infty},\ell_{x,\infty})=\frac{1}{2\pi}(0-n\pi+i\operatorname*{Tr}%
\operatorname{Log}I)+\frac{n}{2}=0
\]
hence formula (\ref{loop}).
\end{proof}

The following generalization of the Maslov index will be used to determine the
effect of a change of initial point on the period orbit. One can extend
definition (\ref{Maslov1}) by associating to every $\ell\in\Lambda(n)$ his own
private Maslov index $\mu_{\ell}$ by the formula
\begin{equation}
\mu_{\ell}(S_{\infty})=m(S_{\infty}\ell_{\infty},\ell_{\infty})
\label{Maslov2}%
\end{equation}
where $\ell_{\infty}$ is the homotopy class in $\Lambda(n)$ of the constant
loop through $\ell$. Obviously formula (\ref{mco}) holds for $\mu_{\ell}$ as
well, replacing $\ell_{p}$ by $\ell$ in the right-hand side. We notice that it
follows from (\ref{cobord}) that
\begin{equation}
\mu_{\ell}(S_{\infty})-\mu_{\ell^{\prime}}(S_{\infty})=\operatorname{Inert}%
(S\ell,\ell,\ell^{\prime})-\operatorname{Inert}(S\ell,S\ell^{\prime}%
,\ell^{\prime})\text{.} \label{dem}%
\end{equation}

Let $S_{0}$ be an arbitrary symplectic matrix and $S_{\infty}\in Sp_{\infty
}(n)$. Then the conjugacy class $S_{0}^{-1}S_{\infty}S_{0}$ is also an element
of $Sp_{\infty}(n)$, and we have:
\begin{equation}
\mu_{\ell}(S_{0}^{-1}S_{\infty}S_{0})=\mu_{S_{0}\ell}(S_{\infty})\text{.}
\label{musli}%
\end{equation}

\section{The Maslov Index of the Monodromy Matrix}

Let $\gamma$ be a periodic orbit of some Hamiltonian $H\in C^{\infty
}(\mathbb{R}_{z}^{2n},\mathbb{R})$; we assume for simplicity that the flow
$(f_{t}^{H})$ determined by $H$ exists for all times. Choose an origin $z$ on
$\gamma$ so that $\gamma(t)=f_{t}^{H}(z)$. The Jacobian matrix $S_{t}%
(z)=Df_{t}^{H}(z)$ then satisfies the \textquotedblleft variational
equation\textquotedblright\
\begin{equation}
\frac{d}{dt}S_{t}(z)=JH^{\prime\prime}(z,t)S_{t}(z)\text{ \ , \ }S_{0}(z)=I
\label{vareq}%
\end{equation}
where $H^{\prime\prime}(z,t)$ is the Hessian matrix of $H$ calculated at
$f_{t}^{H}(z)$. Let $T$ be the prime period of $\gamma$, then $H^{\prime
\prime}(z,t+T)=H^{\prime\prime}(z,t)$. The symplectic matrix $S_{T}(z)$ is
called the \textit{monodromy matrix}, and it satisfies the relation
$S_{t+T}(z)=S_{t}(z)S_{T}(z)$ for all $t$. We denote by $S_{T,\infty}$ the
homotopy class in $Sp(n)$ of the symplectic path $t\longmapsto S_{t}$, $0\leq
t\leq T$. We begin by making a pedestrian remark: since $S_{T}^{r}=S_{rT}$ for
every integer $r\in\mathbb{Z}$ formula (\ref{mco}) allows us at once to
calculate the Maslov index for repetitions of the prime periodic orbit. For
instance, if $r=2$,
\begin{equation}
\mu(S_{2T,\infty})=2\mu(S_{T,\infty})-\operatorname{Inert}(S_{T}^{2}\ell
_{p},S_{T}\ell_{p},\ell_{p})\text{.} \label{two}%
\end{equation}
We will actually use this formula with profit to determine the Maslov index
$\mu(S_{T,\infty})$ itself. Let us namely make the following observation. When
one writes the monodromy matrix $S_{T}$ in exponential form $S_{T}=e^{TX}$ it
is not true in general that the matrix $X$ is in the symplectic Lie algebra;
in fact $X$ is usually not even real! (This is due to the presence of inverse
hyperbolic blocks when $S_{T}$ is put in normal form.) Since $S_{t+2T}%
(z)=S_{t}(z)(S_{T}(z))^{2}$ we can write
\begin{equation}
S_{t}(z)=P_{t}(z)e^{tX(z)} \label{stz}%
\end{equation}
where $t\longmapsto P_{t}(z)$ is $2T$-periodic and $X(z)$ is real; we will
assume that $X(z)\in\mathfrak{sp}(n)$ so that both $P_{t}(z)$ and $e^{tX(z)}$
are symplectic. For notational simplicity we drop for the moment being any
reference to the origin $z$ of the periodic orbit; $z$ will be reinstated when
we set out to discuss the effect of change of origin on the Maslov index.

Let $S_{T},_{\infty}$ (resp. $S_{2T},_{\infty}$) the homotopy class of the
symplectic path $t\longmapsto S_{t}$, $0\leq t\leq T$ (resp. $0\leq t\leq2T$).
Both $S_{T},_{\infty}$ and $S_{2T},_{\infty}$ are elements of $Sp_{\infty}%
(n)$, and $S_{2T},_{\infty}=S_{T}^{2},_{\infty}$.

\begin{theorem}
\label{propoun}$(\ast)$ The Maslov index $\mu(S_{T},_{\infty})$ of the
symplectic path $t\longmapsto S_{t}$, $0\leq t\leq T$, is given by the
formula
\begin{equation}
\mu(S_{T},_{\infty})=\frac{1}{2}(\mu(P_{2T,\infty})+\mu(e_{\infty}%
^{2TX})+\operatorname{Inert}(S_{2T}\ell_{p},S_{T}\ell_{p},\ell_{p}))
\label{ess}%
\end{equation}
where $P_{2T,\infty}$ (resp. $e_{\infty}^{2TX}$) is the homotopy class in
$Sp(n)$ of the path $t\longmapsto P_{t}$ (resp. $t\longmapsto e^{tX}$), $0\leq
t\leq2T$. $(\ast\ast)$ Let $P_{t}=U_{t}e^{Y_{t}}$ be the polar decomposition
of $P_{t}$, that is $U_{t}\in U(n)$ and $Y_{t}$ is a symmetric matrix in
$\mathfrak{sp}(n)$. Then
\begin{equation}
\mu(P_{2T,\infty})=\mu(U_{2T,\infty})=2k \label{wind}%
\end{equation}
where $U_{2T,\infty}$ is the homotopy class in $Sp(n)$ (or in $U(n)$) of
$t\longmapsto U_{t}$ and $k$ is the winding number defined by (\ref{loop}).
$(\ast\ast\ast)$ If the monodromy matrix is $S_{T}=e^{TX}$ then its Maslov
index is given by
\begin{equation}
\mu(S_{T},_{\infty})=\mu(e_{\infty}^{TX})+k \label{sun}%
\end{equation}
where $e_{\infty}^{TX}$ is the homotopy class of $t\longmapsto e^{tX}$, $0\leq
t\leq2T$.
\end{theorem}

\begin{proof}
$(\ast)$ By definition of the group structure of $Sp_{\infty}(n)$ we have
$S_{2T,\infty}=P_{2T,\infty}e_{\infty}^{2TX}$. Using (\ref{mco}), the fact
that $P_{2T}=I$, and the definition of the index of inertia, we have
\begin{align*}
\mu(S_{2T\infty})  &  =\mu(P_{2T,\infty})+\mu(e_{\infty}^{2TX}%
)-\operatorname{Inert}(P_{2T}e^{2TX}\ell_{p},e^{2TX}\ell_{p},\ell_{p})\\
&  =\mu(P_{2T,\infty})+\mu(e_{\infty}^{2TX})-\operatorname{Inert}(e^{2TX}%
\ell_{p},e^{2TX}\ell_{p},\ell_{p})\\
&  =\mu(P_{2T,\infty})+\mu(e_{\infty}^{2TX})
\end{align*}
and formula (\ref{ess}) follows in view of (\ref{two}). $(\ast\ast)$ To prove
(\ref{wind}) we begin by noting that in view of the uniqueness of the
symplectic polar decomposition we have both $U_{0}=U_{2T}=I$ and $e^{Y_{0}%
}=e^{Y_{2T}}=I$. Writing $P_{2T,\infty}=U_{2T,\infty}e_{\infty}^{Y_{2T}}$ we
have, again by (\ref{mco}) and the definition of the index of inertia:
\begin{align*}
\mu(P_{2T,\infty})  &  =\mu(U_{2T,\infty})+\mu(e_{\infty}^{Y_{2T}%
})-\operatorname{Inert}(P_{2T}\ell_{p},e^{Y_{2T}}\ell_{p},\ell_{p})\\
&  =\mu(U_{2T,\infty})+\mu(e_{\infty}^{Y_{2T}})
\end{align*}
because $P_{2T}\ell_{p}=e^{Y_{2T}}\ell_{p}=\ell_{p}$ implies that%
\[
\operatorname{Inert}(P_{2T}\ell_{p},e^{Y_{2T}}\ell_{p},\ell_{p}%
)=\operatorname{Inert}(\ell_{p},\ell_{p},\ell_{p})=0\text{.}%
\]
Formula (\ref{wind}) will follow if we show that $\mu(e_{\infty}^{Y_{2T}})=0$.
Now, $e_{\infty}^{Y_{2T}}$ is the homotopy class in $Sp(n)$ of the loop
$t\mapsto e^{Y_{t}}$, $0\leq t\leq T$. The subset of $Sp(n)$ consisting of
positive definite matrices is simply connected, hence that loop is
contractible to a point, and thus homotopic to the identity. The relation
$\mu(e_{\infty}^{Y_{2T}})=0$ follows in view of (\ref{propelo}). $(\ast
\ast\ast)$ By definition of the product in $Sp_{\infty}(n)$ we have
$e_{\infty}^{2TX}=e_{\infty}^{TX}e_{\infty}^{TX}$ and hence
\[
\mu(e_{\infty}^{2TX})=2\mu(e_{\infty}^{TX})-\operatorname{Inert}(e^{2TX}%
\ell_{p},e^{TX}\ell_{p},\ell_{p})
\]
in view of (\ref{mco}). It follows from (\ref{ess}) and (\ref{wind}) that
\begin{align*}
\mu(S_{T},_{\infty})  &  =\mu(e_{\infty}^{TX})+k+\\
&  \frac{1}{2}(\operatorname{Inert}(S_{2T}\ell_{p},S_{T}\ell_{p},\ell
_{p})-\operatorname{Inert}(e^{2TX}\ell_{p},e^{TX}\ell_{p},\ell_{p}))
\end{align*}
hence formula (\ref{sun}) since $e^{2TX}=S_{2T}$ and we are assuming that
$e^{TX}=S_{T}$.
\end{proof}

There remains to investigate what happens to the Maslov index when we change
the initial point $z$. Let us consider two solutions $t\longmapsto S_{t}(z)$
and $t\longmapsto S_{t}(z^{\prime})$ of the variational equation (\ref{vareq})
where $z$ and $z^{\prime}$ are two points on the same periodic orbit $\gamma$
of the Hamiltonian $H$. We begin with the following preliminary remark: since
$z$ and $z^{\prime}$ are on the same orbit there exists a time $t^{\prime}$
such that $z=f_{t^{\prime}}^{H}(z^{\prime})$ and hence, using the chain rule
together with the relation $f_{t+t^{\prime}}^{H}=f_{t}^{H}f_{t^{\prime}}^{H}%
$:
\begin{equation}
S_{t+t^{\prime}}(z^{\prime})=S_{t}(z)S_{t^{\prime}}(z^{\prime}) \label{stt}%
\end{equation}
(recall that $S_{t}(z)$ is the Jacobian matrix of $f_{t}^{H}$ calculated at
$z$).

\begin{theorem}
$(\ast)$ The Maslov indices $\mu(S_{T},_{\infty}(z))$ and $\mu(S_{T},_{\infty
}(z^{\prime}))$ corresponding to two points $z$ and $z^{\prime}$ lying on the
same periodic orbit of the Hamiltonian $H$ are related by the formula:
\begin{equation}
\mu(S_{T},_{\infty}(z^{\prime}))=\mu_{S^{\prime}\ell_{p}}(S_{T},_{\infty
}(z))\text{ \ \textrm{with} \ }S^{\prime}=S_{t^{\prime}}(z^{\prime})\text{.}
\label{u1}%
\end{equation}
Equivalently, in view of (\ref{dem}):
\begin{equation}
\mu(S_{T},_{\infty}(z))-\mu(S_{T},_{\infty}(z^{\prime}))=\operatorname{Inert}%
(S_{T}\ell_{p},\ell_{p},S^{\prime}\ell_{p})-\operatorname{Inert}(S_{T}\ell
_{p},S_{T}S^{\prime}\ell_{p},S^{\prime}\ell_{p})\text{.} \label{u2}%
\end{equation}
$(\ast\ast)$ The winding number $\mu(U_{2T,\infty})$ in (\ref{wind}) is not
affected by the change of $z$ in $z^{\prime}$:
\begin{equation}
\mu(U_{2T},_{\infty}(z))=\mu(U_{2T},_{\infty}(z^{\prime}))\text{.} \label{mut}%
\end{equation}

\end{theorem}

\begin{proof}
$(\ast)$ Formulas (\ref{u1}) and (\ref{u2}) are equivalent in view of
(\ref{dem}). It follows from (\ref{stt}) that we have $S_{t^{\prime}%
+T}(z^{\prime})=S_{T}(z)S^{\prime}$; since on the other hand $S_{t^{\prime}%
+T}(z^{\prime})=S_{t^{\prime}}(z^{\prime})S_{T}(z^{\prime}) $ by the
properties of the monodromy matrix, we thus have the following conjugacy
relation between the monodromy matrices at different points of the orbit:
\begin{equation}
S_{T}(z^{\prime})=(S^{\prime})^{-1}S_{T}(z)S^{\prime}\text{.} \label{conjug}%
\end{equation}
We next note that the two symplectic paths
\[
\Sigma:t\longmapsto S_{t}(z^{\prime})\text{ \ \textrm{and} \ }\Sigma^{\prime
}:t\longmapsto(S^{\prime})^{-1}S_{t}(z)S^{\prime}\text{ \ }(0\leq t\leq T)
\]
are homotopic with fixed endpoints (that they are homotopic is clear letting
$t^{\prime}\rightarrow0$; they moreover have same endpoints $I$ and
$S_{t}(z^{\prime})$ in view of (\ref{conjug})). These paths thus have same
Maslov index $\mu(S_{T},_{\infty}(z^{\prime}))$, and hence
\[
\mu(S_{T},_{\infty}(z^{\prime}))=\mu((S^{\prime})^{-1}S_{T,\infty}%
(z)S^{\prime})=\mu_{S^{\prime}\ell_{p}}(S_{T,\infty}(z))
\]
proving (\ref{u1}). $(\ast\ast)$ We have, in view of (\ref{dem}), and using
the fact that $U_{T}=I$:
\begin{align*}
\mu(U_{2T},_{\infty}(z))-\mu(U_{2T},_{\infty}(z^{\prime}))  &
=\operatorname{Inert}(U_{2T}\ell_{p},\ell_{p},S^{\prime}\ell_{p}%
)-\operatorname{Inert}(U_{2T}\ell_{p},U_{T}S^{\prime}\ell_{p},S^{\prime}%
\ell_{p})\\
&  =\operatorname{Inert}(\ell_{p},\ell_{p},S^{\prime}\ell_{p}%
)-\operatorname{Inert}(\ell_{p},S^{\prime}\ell_{p},S^{\prime}\ell_{p})\\
&  =0\text{.}%
\end{align*}

\end{proof}

\section{Relation With Gutzwiller's Index}

Let us shortly explain in which way the Maslov index we have constructed is
related to the index appearing in Gutzwiller's trace formula%
\[
\delta g(E)=\frac{1}{\pi\hbar}\sum_{\gamma}\frac{T_{\gamma}}{\sqrt
{|\det(\tilde{M}_{\gamma}-I)}}\cos(\frac{1}{\hbar}\mathcal{A}_{\gamma}%
-\frac{\pi}{2}\xi_{\gamma})
\]
for the oscillating part of the level density of a system which has only
isolated periodic orbits $\gamma$ with prime periods $T_{\gamma}$.\ In the
formula above $\tilde{M}_{\gamma}$ is the stability matrix of the periodic
orbit $\gamma$, and $\mathcal{A}_{\gamma}$ the action of that orbit; the term
$\frac{\pi}{2}\xi_{\gamma}$ can be viewed as giving the argument of the square
root $\sqrt{|\det(\tilde{M}_{\gamma}-I)}$. One shows \cite{CL,CRL,Robbins}
that $\xi_{\gamma}$ can be written as a sum
\[
\xi_{\gamma}=\mu_{\gamma}+\nu_{\gamma}%
\]
of two contributions: $\mu_{\gamma}$ is the phase index appearing in the
semi-classical time-dependent Green function $G(x,x^{\prime};t)$ while
$\nu_{\gamma}$ arises when taking the trace of $G$ by stationary phase
integrations transverse to $\gamma$ (see \cite{Gutz} for a detailed account of
these procedures); it is shown in \cite{CRL} that $\xi_{\gamma}$ does not
depend on the choice of the initial point on the orbit $\gamma$ while the
individual terms $\mu_{\gamma}$, $\nu_{\gamma}$ are in general dependent of
that point. We can give the following (tentative) interpretation of the
relation between our constructs and the Gutzwiller index $\xi_{\gamma}$.
Assume that the orbit $\gamma$ lies on an invariant Lagrangian submanifold
$\mathbb{V}$ of phase space. To each point $\gamma(t) $ of that orbit we can
associate a Lagrangian plane $\ell(t)$, namely the tangent space to
$\mathbb{V}$ at $\gamma(t)$. We thus obtain a loop $\lambda$ in the Lagrangian
Grassmannian $\Lambda(n)$. Now there is a fundamental relation between the
Leray index and the usual Arnol'd--Maslov index $\mathrm{Mas}(\lambda)$ of the
loop $\lambda$: we have, for every $\ell_{\infty}\in\Lambda_{\infty}(n)$,
\begin{equation}
\mathrm{Mas}(\lambda)=m(\ell_{\infty}(T),\ell_{\infty})-m(\ell_{\infty
}(0),\ell_{\infty}) \label{mas}%
\end{equation}
where $\ell_{\infty}(0)$ is the homotopy class in $\Lambda(n)$ of the origin
$\ell(0)=\lambda(0)$ of $\lambda$ and $\ell_{\infty}(T)$ that of the whole
loop (see \cite{JMPA,Wiley,Leray}). We now observe that since $\gamma(t)$ is
obtained from $\gamma(t)$ by $f_{t}^{H}$ it follows that $\lambda(t)$ is
obtained from $\lambda(0)$ by $S_{t}$; choosing $\ell_{\infty}=\ell_{\infty
}(0)$ in (\ref{mas}) and using the first equality (\ref{propel}), we get%
\[
\mathrm{Mas}(\lambda)=m(S_{T,\infty}\ell_{\infty}(0),\ell_{\infty}%
(0))=\mu_{\ell(0)}(S_{T,\infty})\text{.}%
\]
If --as suggested in \cite{CRL} and \cite{Robbins}-- we have $\xi_{\gamma
}=\mathrm{Mas}(\lambda)$, it follows that we actually have
\begin{equation}
\xi_{\gamma}=\mu_{\ell(0)}(S_{T,\infty})\text{.} \label{kmu}%
\end{equation}
Note that since $\mathrm{Mas}(\lambda)$ is independent of the choice of origin
so is the case for $\mu_{\ell(0)}(S_{T,\infty})$ (this property also easily
follows from (\ref{musli}) and (\ref{conjug}), alternatively (\ref{u2})).

We will come back to a detailed analysis of the relation between the
Gutzwiller and Leray indices (and in particular of the validity of
(\ref{kmu})) in a forthcoming paper.

\end{document}